\documentclass[preprint]{revtex4}
\usepackage{graphicx}

\begin{document}

\markboth{Seoktae Koh}
{Trans-Planckian Physics and Non-Commutative Inflation}


\title{Trans-Planckian Physics and Non-Commutative Inflation}

\author{Seoktae Koh}
\address{Department of Physics, Hanyang University, Seoul 133-791, Korea \\
and Institute of Theoretical Physics, Chinese Academy of Sciences, \\
P.O. Box 2735, Beijing, 100080, China \\
skoh@itp.ac.cn}



\begin{abstract}
Non-commutativity of spacetime at the Planck scale may deform the
usual dispersion relations. And these deformed dispersion
relations could lead to the accelerating phase without a scalar field.
In this paper, we have calculated the spectral index
and the running of spectral index in a non-commutative inflation model. 
Non-commutative
inflation with thermal radiation 
gives a scale invariant spectrum in the limit $w \rightarrow
-1$ and negative running spectral index which are consistent with
the WMAP 3-year results.

\end{abstract}

\maketitle

\section{Introduction} 
It is realized from the fundamental theory ({\it e.g.} string theory)
 that spacetime structure will become non-commutative at the Planck scale.
And it is also believed that inflation can provide a promising framework 
to probe Planck scale physics through the prediction of cosmological
perturbations which arise due to the non-commutativity. 
There are several attempts to show the effect of the non-commutativity
of the spacetime on the cosmological observations \cite{Alexander01b}\cite{Brandenberger02}\cite{Kim05}. 

Spacetime non-commutativity may result in deformed dispersion relations
\cite{Alexander01a}.
It was pointed out in Ref.~\cite{Alexander01b} that these
 deformed dispersion
relations lead to a period of inflation driven by non-commutative 
radiation instead of usual scalar fields. 
Its distinct features are thermal fluctuations as seeds for structure
formation and the existence of 
two branches in the deformed dispersion relation. High energy branch
,in which the energy decrease as the momentum increases, leads to an
accelerated expansion phase.
The resulting spectrum of
the cosmological perturbations in this model produced a scale invariant
spectrum in the limit $w\rightarrow -1$~\cite{Koh07a}.

The following section provides a brief review of the non-commutative inflation
model and section \ref{cosmo_pertb} gives a detail calculation of
the power spectrum in this model. In section \ref{observ}, we
calculate a spectral index and a running of spectral index and then compare
with the observational data. And finally we 
conclude with a brief discussion.

\section{Brief Review of Non-Commutative Inflation}
The modified dispersion relations for  massless particles
which  result from the 
space-time non-commutativity,
\begin{eqnarray}
E^2- p^2 c^2 f(E)^2 = 0
\end{eqnarray}
 drive the accelerated expansion of space. Here,
\begin{eqnarray}
f(E) = 1+(\lambda E)^{\alpha},
\end{eqnarray}
where $p$ and $E$ denote momentum and energy, respectively, 
$\alpha \geq 1$ is a positive constant, and $\lambda$, 
which determines the maximal momentum, has a length scale.

\begin{figure}[t]
\includegraphics[width=2.5in ]{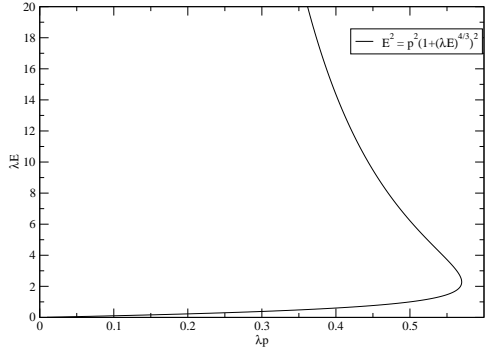}
\caption{Modified dispersion relations due to the non-commutativity of
spacetime for $\alpha = 4/3$. While for a low energy branch 
energy increases as momentum increases,
 as momentum increases, energy decrease in a high energy branch.
\protect\label{dp}}
\end{figure}


These dispersion relations have two branches for $\alpha >1$
(see Fig. \ref{dp}). For the high energy branch, the energy
increases as the momentum decreases. This behavior can make it
possible to generate accelerated expansion phase.

The energy density with the modification of the dispersion relations
is given by 
\begin{eqnarray}
\rho = \frac{1}{\pi^2}\int dE\frac{E^3}{e^{E/T}-1}\frac{1}{f^3}
\left|1-\frac{f^{\prime}E}{f}\right|,
\end{eqnarray}
and the pressure is 
\begin{eqnarray}
{\cal P} = \frac{1}{3\pi^2}\int \frac{dE}{1-\frac{f^{\prime}E}{f}}
\frac{E^3}{e^{E/T}-1}\frac{1}{f^3}
\left|1-\frac{f^{\prime}E}{f}\right|,
\end{eqnarray}
where $f^{\prime} \equiv df/dE$.
 While for 
low temperatures $\rho \propto T^4$ as usual radiation does,
 $\rho \propto T$ for $\lambda T \gg 1$\cite{Alexander01a}.
In the high energy limit ($\lambda E \gg 1$), $f^{\prime}E/f \simeq 
\alpha$ and the equation of state parameter $w$ takes on negative 
values,
\begin{eqnarray}
w \simeq \frac{1}{3(1-\alpha)}
\end{eqnarray}
As temperature drops, the equation of state of non-commutative radiation
changes from that of an accelerated universe to $\frac{1}{3}$
which is an equation of state of
ordinary radiation. Thus, there is no need for any period of  reheating.

The Friedmann and energy conservation equations for the non-commutative
radiation are
\begin{eqnarray}
\left(\frac{\dot{a}}{a}\right)^2 &=& \frac{8\pi}{3M_p}\rho, \nonumber \\
\dot{\rho} &=& -3 \frac{\dot{a}}{a}(1+w)\rho.
\end{eqnarray}
For a constant equation of state, the background solutions can be written as
\begin{eqnarray}
\rho \propto a^{-3(1+w)}, \quad a\propto t^{2/3(1+w)} \propto
\eta^{2/(1+3w)},
\label{bgsol}
\end{eqnarray}
where $\eta$ is a conformal time ($dt = a d\eta$).
Inflation ends when there are no more excited states on the high energy
branch of the modified dispersion relations.

\section{Density Perturbations in Non-Commutative Inflation} \label{cosmo_pertb}
We work in the longitudinal gauge to calculate the density perturbations
of non-commutative fluid. In this gauge, the metric takes the form
\begin{eqnarray}
ds^2 = a^2[-(1+2\Phi) +(1-2\Psi)\gamma_{ij}dx^i dx^j]
\end{eqnarray}
where the functions $\Phi$ (the Bardeen potential\cite{Bardeen80})
 and $\Psi$ describe the scalar metric fluctuations.

Using the variable $v$ which is defined as\cite{Mukhanov90}
\begin{eqnarray}
v = \frac{a}{c_s\sqrt{\rho+{\cal P}}}\left(\delta q-\frac{\rho+{\cal P}}{H}
\Phi\right) = -z \zeta, \quad 
z = \frac{a \sqrt{\rho+{\cal P}}}{H c_s}
\label{sm}
\end{eqnarray}
the equation of motion for scalar metric perturbations takes the simple 
form
\begin{eqnarray}
v_k^{\prime\prime} +\left(k^2 c_s^2-\frac{z^{\prime\prime}}{z}\right)v_k
= 0,
\end{eqnarray}
where $c_s^2 \equiv \delta {\cal P}/\delta \rho$ is a sound speed,
 $\delta q$ a velocity perturbation and a prime denotes a derivative
with respect to conformal time $\eta$. The curvature perturbation 
$\zeta$ on uniform energy density hypersurface, which
is conserved on scales larger than the sound horizon for
purely adiabatic perturbations, is related to $\Phi$ as
\begin{eqnarray}
\zeta = \Phi - \frac{H}{\dot{H}}(\dot{\Phi}+H\Phi).
\end{eqnarray}

Then the power spectrum of $\zeta$ is given by
\begin{eqnarray}
{\cal P}_{\zeta} (k) = \frac{k^3}{2\pi^2}\langle \theta|
\zeta^{\ast}_k \zeta_k|\theta \rangle 
=\frac{k^3}{2\pi^2}\frac{\langle \theta| v^{\ast}_k v_k|\theta \rangle}
{z^2}.
\label{ps}
\end{eqnarray}

The seeds for large scale structures in this non-commutative inflation
model are thermal fluctuations instead of quantum fluctuations which 
are responsible for structure formation in the conventional scalar
driven inflation models. We need initial conditions 
to calculate completely
$v_k$ or curvature perturbations $\zeta_k$ at superhorizon scales. 
 For a fixed scale $k$,
 we choose  thermal correlation length to impose 
initial conditions\cite{Koh07a}
\begin{eqnarray}
T^{-1}(\eta_i(k)) = a(\eta_i(k))/k,
\end{eqnarray}
since on scale larger than that the interactions are not likely able
to maintain thermal equilibrium.
Thermal state $|\theta \rangle$ at time $\eta_i(k)$ is
determined by $n_k = \langle \theta |a_k^{\dag}
a_k |\theta\rangle = 1/(e^{E/aT}-1)$.

In the Heisenberg picture, the Hamiltonian $H_k$ is given by
\begin{eqnarray}
H_k = \int d^3 k\frac{1}{2}\left[k(a_k a_k^{\dag} + a_k^{\dag}a_k)
+i \frac{z_k^{\prime}}{z_k}(a_k^{\dag}a_{-k}^{\dag}
-a_k a_{-k})\right].
\end{eqnarray}
The annihilation $a_k$ and creation operator $a_k^{\dag}$ at time $\eta$
are related to those at the initial time $\eta_i$ by the Bogoliubov
transformation
\begin{eqnarray}
a_k(\eta) &=& \alpha_k(\eta)a_k(\eta_i) +\beta_k(\eta) a_{-k}^{\dag}(\eta_i),
\nonumber \\
a_{-k}^{\dag}(\eta) &=& \alpha_k^{\ast}(\eta) a_{-k}^{\dag}(\eta_i)
+\beta_{k}^{\ast}(\eta)a_k(\eta_i), 
\end{eqnarray}
where $\alpha_k$ and $\beta_k$ are the Bogoliubov coefficients which 
satisfy the normalization condition $|\alpha_k|^2 - |\beta_k|^2 = 1$. The
canonical variable $v_k$ and its conjugate momenta $\pi_k$ yields
\begin{eqnarray}
v_k (\eta) &=& f_k(\eta) a_k(\eta_i) +f_k^{\ast}a_{-k}^{\dag}(\eta_i), 
\nonumber \\
\pi_k(\eta) &=& -i [g_k(\eta) a_k(\eta_i) +g_k^{\ast}(\eta)a_{-k}^{\dag}
(\eta_i)],
\label{mod_exp}
\end{eqnarray}
where 
\begin{eqnarray}
f_k(\eta) &=& \frac{1}{\sqrt{2k c_s}}(\alpha_k(\eta) +
\beta_k^{\ast}(\eta)), \nonumber \\
g_k(\eta) &=& \sqrt{\frac{kc_s}{2}}(\alpha_k(\eta)-\beta_k^{\ast}(\eta)).
\end{eqnarray}

For a power law inflation with a constant $w$, the mode equation has the
following solution
\begin{eqnarray}
f_k(\eta) = A_k\sqrt{-\eta}J_{\nu}(-kc_s\eta)
+B_k\sqrt{-\eta}Y_{\nu}(-kc_s\eta),
\label{mod_sol}
\end{eqnarray}
where $J_{\nu}$ and $Y_{\nu}$ are Bessel functions of the first
and second kind, respectively, and we have used 
\begin{eqnarray}
\frac{z^{\prime\prime}}{z} = \frac{\nu^2-1/4}{\eta^2}, \quad
\nu^2 = \frac{1}{4} +\frac{2(1-3w)}{(1+3w)^2}.
\label{def_nu}
\end{eqnarray}
The expression of $A_k$ and $B_k$ can be obtained from 
the normalization condition for $\alpha_k$ and $\beta_k$ and
$\beta_k^{\ast}(\eta_i) = 0$ at the initial time $\eta_i$
\begin{eqnarray}
|A_k|^2 &=& -\frac{\pi^2}{8} kc_s\eta_i
(Y_{\nu}^2(-kc_s\eta_i)+Y_{\nu-}^2(-kc_s\eta_i)),\nonumber \\
|B_k|^2 &=& -\frac{\pi^2}{8} kc_s\eta_i
(J_{\nu}^2(-kc_s\eta_i)+J_{\nu-1}^2(-kc_s\eta_i)),\nonumber \\
A_kB_k^{\ast} &=& -i\frac{\pi}{4}
+\frac{\pi^2}{8} k c_s \eta_i (J_{\nu}(-kc_s\eta_i)
Y_{\nu}(-kc_s\eta_i) \nonumber \\
& &~~~~~~~~~~~~~~~~~~+J_{\nu-1}(-kc_s\eta_i)
Y_{\nu-1}(-kc_s\eta_i)).
\end{eqnarray}

In the large scale limit, $k c_s\eta \ll 1$, 
Bessel functions have asymptotic form as
\begin{eqnarray}
J_{\nu}(-kc_s \eta) \simeq \frac{1}{\Gamma(\nu+1)}
\left(-\frac{kc_s\eta}{2}\right)^{\nu}, \quad
Y_{\nu}(-kc_s\eta) \simeq -\frac{\Gamma(\nu)}{\pi}
\left(-\frac{kc_s\eta}{2}\right)^{-\nu},
\end{eqnarray}
where $\nu = 3(\omega-1)/2(1+3\omega)$ from (\ref{def_nu}).
Then curvature perturbation in the large scale limit
can be written as
\begin{eqnarray}
\zeta_k(\eta) \simeq c_k +d_k \eta^{-\frac{3(1-w)}{1+3w}},
\end{eqnarray}
where $c_k$ and $d_k$ are growing and decaying mode coefficients, 
respectively. As expected, $\zeta_k$ has a constant solution in the 
large scale limit \cite{Koh07b}.

Using  (\ref{mod_exp}) and (\ref{mod_sol}), 
the power spectrum (\ref{ps}) of  $\zeta_k$ 
at the time of Hubble radius crossing
can now be calculated as 
\begin{eqnarray}
{\cal P}_{\zeta}(k) &=& 
 \frac{k^3}{2\pi^2}\frac{|f_k|^2}{z^2}\coth\left(\frac{kc_s}{a_i T_i}
\right) \nonumber \\
&\sim& \frac{2\pi}{3M_p^4}
\frac{k^2 c_s^2 \rho(\eta_h)\eta_i \eta_h}{\epsilon c_s} \nonumber \\
&\sim& \frac{2\pi^2}{3M_p^4}\frac{\rho(\eta_h)}{\epsilon}\frac{T_i}{c_s H_i},
\label{pspectrum}
\end{eqnarray}
where we have used $z \simeq M_p a\epsilon^{1/2}/\sqrt{4\pi c_s^2}$ and
 $\eta_i \simeq 1/(a_i H_i)$  and $\eta_h \simeq 1/(a_i T_i)$ since
$kc_s = a_i T_i = a(\eta_h) H(\eta_h)$.
For later use, we have defined slow-roll parameters
\begin{eqnarray}
\epsilon \equiv -\frac{\dot{H}}{H^2}, 
\quad \eta \equiv \frac{\ddot{H}}{H\dot{H}},
\quad \xi \equiv \frac{H^{(3)}}{H^2 \dot{H}},
\end{eqnarray}
where $H^{(3)} \equiv d^3 H/dt^3$.

\section{Observations} \label{observ}
One important feature of the thermal 
non-commutative inflation model is the existence of two branches in
dispersion relations. This can make it possible to generate
 accelerated  expansion phase and, after a period of inflation, change 
into the usual radiation dominated phase smoothly.
In order to differentiate between non-commutative inflation model
and usual scalar field-driven inflation model, it is necessary
to constrain the power spectrum of the cosmological perturbations
\begin{eqnarray}
\ln {\cal P}_{\zeta}(k) = \ln {\cal P}_{\zeta}(k_0)
+(n_s-1)\ln\frac{k}{k_0} +\frac{1}{2}\alpha_s \ln^2 \frac{k}{k_0}+
\cdots,
\end{eqnarray}
where $n_s$ is a spectral index of the spectrum and $\alpha_s$ 
a running of the spectral index, respectively. 
Three-year WMAP results\cite{Spergel06} show that 
\begin{eqnarray}
n_s = 1.16 \pm 0.10,\quad \alpha_s = -0.085 \pm 0.043
\end{eqnarray}
on the scale $k_0 = 0.002{\rm Mpc}^{-1}$
\footnote{If a running of spectral index is not taken into account,
the best fit values\cite{Spergel06} 
of the spectral index is $n_s = 0.958\pm 0.016$.
On the contrary, if a running spectral index is included, 
while $n_s > 1$ is favored on the scale $k_0 = 0.002{\rm Mpc}^{-1}$,
$n_s < 1$ is favored at $k_0 = 0.05{\rm Mpc}^{-1}$.}.

Here we briefly
mention the observational implications of the thermal non-commutative
inflation model.  
It is convenient to use the following relations to calculate the spectral 
index and its running:
\begin{eqnarray}
\frac{d\ln \rho}{d\ln k} = -2\epsilon, \quad
\frac{d\epsilon}{d\ln k } = 2\epsilon^2 - \epsilon \eta, \quad
\frac{d\eta}{d\ln k} = \xi + \epsilon \eta - \eta^2, 
\label{dslowroll}
\end{eqnarray}
where we have used $d\ln k \simeq H dt$ when the modes cross
the Hubble radius.
Then the spectral index from (\ref{pspectrum}) becomes
\begin{eqnarray}
n_s -1 \equiv \frac{d\ln {\cal P}_{\zeta}(k)}{d\ln k} 
= -4\epsilon +\eta +\frac{d}{d\ln k}\ln\left(\frac{T_i}{c_s H_i}\right).
\label{spec_index}
\end{eqnarray}
For a constant equation of state ($\dot{w}\ll w$), using
Eq. (\ref{bgsol})
 slow-roll parameters become 
\begin{eqnarray}
& &\epsilon \simeq \frac{3}{2}(1+w), \quad
\eta \simeq -3(1+w), \quad \xi \simeq \frac{27}{2}(1+w)^2, 
\label{slowroll2}\\
& &\frac{d}{d\ln k}\ln \left(\frac{T_i}{c_s H_i}\right) \sim
\frac{3(1+w)}{2(2+3w)},
\label{corr}
\end{eqnarray}
where we have used $\rho \propto T$ during inflation to derive 
Eq. (\ref{corr}).
Then the spectral index (\ref{spec_index}) becomes 
\begin{eqnarray}
n_s - 1 \simeq -\frac{3(1+w)(18w+11)}{2(2+3w)}.
\end{eqnarray}
This gives a scale invariant spectrum in the limit $w\rightarrow -1$ with
a slightly red tilt\cite{Koh07a}\footnote{More exact form of
the spectral index is given in Ref. ~\cite{Koh07a} using $\rho
\propto a^{-3(1+w)}\propto \eta^{-\frac{6(1+w)}{1+3w}}$ at the
horizon crossing ($kc_s \eta = 1$). In this paper, we have neglected the
variations of $H$ during inflation period.}. It seems to be inconsistent
with the WMAP three-year data which favor $n_s > 1$ on the large scales.

Using the relations ~(\ref{dslowroll}),
the running of spectral index $\alpha_s$ becomes
\begin{eqnarray}
\alpha_s &\equiv& \frac{d n_s}{d\ln k} = -8 \epsilon^2
+5\epsilon \eta -\eta^2 +\xi +\frac{d^2}{d\ln k^2}
\ln\left(\frac{T_i}{c_s H_i}\right) \\
&\simeq& -36 (1+w)^2,
\end{eqnarray}
where we have used a constant equation of state to derive the second line
in the above expression.
This gives a negative value of running of spectral index
 which is consistent with the observational data.

Another important observational signature is a non-Gaussianity of
cosmological perturbations\cite{Maldacena02}. 
Recent study\cite{Yadav07} shows that 
non-Gaussianity at $95\%$ confidence level is in the region
\begin{eqnarray}
26.91 < f_{NL} <146.71,
\end{eqnarray}
where non-Gaussianity parameter $f_{NL}$ is defined as
\begin{eqnarray}
\zeta = \zeta_g +\frac{3}{5}f_{NL}(\zeta_g^2 - \langle \zeta_g^2\rangle).
\end{eqnarray}
Here,  subscript $g$ denotes the Gaussian part of $\zeta$.
Detection of non-Gaussianity from the
future satellite experiment
 can discriminate between different inflation models.

Interestingly, in Ref. ~\cite{Chen07}, it was shown
 that thermal fluctuations
may lead to a large non-Gaussianity. The non-Gaussianity parameter
takes the form in the thermal fluctuation model 
\begin{eqnarray}
f_{NL} =\frac{5\epsilon m}{72\pi G m\rho L^2},
\end{eqnarray}
where $\rho \propto T^m$ and $L$ is a thermal length scale. This result
implies that as $L$ is smaller, $f_{NL}$ can be larger. 
The non-Gaussianity in thermal non-commutative inflation model is
under considerations.\cite{Koh08b}

\section{Conclusion and Discussions}
In this paper we have considered the inflation model which arises
due to the non-commutativity of spacetime in the early universe and
 calculated the spectral index and
the running of spectral index  in this model. 
Non-commutativity of spacetime at the Planck 
scale may deform usual dispersion relations. Further,
these modified dispersion relations make it possible to 
obtain a period of inflation driven by non-commutative radiation.
While the seeds for large scale structure in the usual scalar
field driven inflation model are quantum fluctuations, those are 
thermal fluctuations in non-commutative inflation. 
Keeping these in mind, it seems natural to impose the initial conditions
at thermal correlation length to maintain thermal equilibrium.
In Ref.~\cite{Koh07a}, it was shown that
this inflation model can give a scale invariant
spectrum in the limit $w\rightarrow -1$.

One of the important features of this model is the existence of  two
branches. In the high energy limit, energy decreases as momentum increases.
This property is crucial in realizing accelerated expansion phase in
modified dispersion relations.
On the contrary, as momentum decreases, energy decreases in the low energy
limit as in usual dispersion relations. Thus, as temperature drops,
the accelerated expansion phase change into the usual radiation dominated
phase ($w = 1/3$) smoothly without need of reheating.

It might be  worth to find the observational signal of these
distinct features of non-commutative inflation model 
to discriminate from the usual scalar field driven inflation model. 
We thus have calculated the spectral index and the running of spectral index.
With the slow-roll approximation, $n_s - 1 \sim -9 (1+w)$ and
$\alpha_s \sim -36 (1+w)^2$. These show that power spectrum is
scale invariant 
as long as $\epsilon, \eta \ll 1$(or $w\rightarrow -1$),
 which is a necessary condition for occurring enough 
inflation, and has negative running of spectral index. While the running
of spectral index is consistent with data, the spectral index seems to 
be inconsistent since WMAP three-year data favor $n_s > 1$.

Another important observational signal is a non-Gaussianity of cosmological
perturbations whose detection will be an important mission in the future
satellite experiment.  Thermal
fluctuations can lead to a large non-Gaussianity\cite{Chen07} \cite{Chen07b}
if thermal length scale is smaller than Horizon scale at the horizon crossing. 
The non-Gaussianity
in this non-commutative inflation model is under considerations\cite{Koh08b}. 

These observational
signals can provide the possibility to probe Planck scale physics in the 
current observations by the induced signals in the spectrum
of cosmological perturbations. 

\section*{Acknowledgments}

The author would like to thank the organizers for 
inviting me and their warm hospitality
during CosPA 2007 at National Taiwan University. And the author also
thank R. Brandenberger and R. Cai for useful comments.
 This work was supported
 by the Korea Science and Engineering Foundation (KOSEF)
(No. R01-2006-000-10651-0).


\begin{thebibliography}{0}


\bibitem{Alexander01b}
 S.~Alexander, R.~Brandenberger and J.~Magueijo,
 Phys.\ Rev.\ D {\bf 67}, 081301 (2003)
 [arXiv:hep-th/0108190].

\bibitem{Brandenberger02}
 R.~Brandenberger and P.~M.~Ho,
 Phys.\ Rev.\ D {\bf 66}, 023517 (2002)
 [AAPPS Bull.\ {\bf 12N1}, 10 (2002)]
 [arXiv:hep-th/0203119].

\bibitem{Kim05}
 H.~C.~Kim, J.~H.~Yee and C.~Rim,
 Phys.\ Rev.\ D {\bf 72}, 103523 (2005)
 [arXiv:gr-qc/0506122].

\bibitem{Alexander01a}
 S.~Alexander and J.~Magueijo,
 arXiv:hep-th/0104093.

\bibitem{Koh07a}
 S.~Koh and R.~H.~Brandenberger,
 JCAP {\bf 0706}, 021 (2007)
 [arXiv:hep-th/0702217].

\bibitem{Bardeen80}
  J.~M.~Bardeen,
  Phys.\ Rev.\  D {\bf 22}, 1882 (1980).

\bibitem{Mukhanov90}
  V.~F.~Mukhanov, H.~A.~Feldman and R.~H.~Brandenberger,
  Phys.\ Rept.\  {\bf 215}, 203 (1992).

\bibitem{Koh07b}
 S.~Koh and R.~H.~Brandenberger,
 JCAP {\bf 0711}, 013 (2007)
 [arXiv:0708.1014 [hep-th]].

\bibitem{Spergel06}
 D.~N.~Spergel {\it et al.} [WMAP Collaboration],
 Astrophys.\ J.\ Suppl.\ {\bf 170}, 377 (2007)
 [arXiv:astro-ph/0603449].

\bibitem{Maldacena02}
  J.~M.~Maldacena,
  JHEP {\bf 0305}, 013 (2003)
  [arXiv:astro-ph/0210603];
  N.~Bartolo, E.~Komatsu, S.~Matarrese and A.~Riotto,
  Phys.\ Rept.\  {\bf 402}, 103 (2004)
  [arXiv:astro-ph/0406398].

\bibitem{Yadav07}
 A.~P.~S.~Yadav and B.~D.~Wandelt,
 arXiv:0712.1148 [astro-ph].

\bibitem{Chen07}
 B.~Chen, Y.~Wang and W.~Xue,
 arXiv:0712.2345 [hep-th].


\bibitem{Koh08b} R. Brandenberger, B. Chen, and S. Koh, in preparation.


\bibitem{Chen07b}
  B.~Chen, Y.~Wang, W.~Xue and R.~Brandenberger,
  arXiv:0712.2477 [hep-th].

\end{thebibliography}
\end{document}